\documentclass[pre,aps,twocolumn,showpacs,floatfix,superscriptaddress]{revtex4}
\usepackage{epsfig}
\usepackage{bm}
\usepackage{times}
\usepackage{mathptm}
\usepackage{mathptmx}
\usepackage{amsmath}
\usepackage{graphicx}
\usepackage{graphics}

\begin{document}
\title{Random graph model with power-law distributed triangle subgraphs}
\author{Danilo Sergi}
\affiliation{D\'epartement de Physique Th\'eorique,
Universit\'e de Gen\`eve, 1211 Gen\`eve 4,
Switzerland}
\date{\today}
\pacs{89.75.Da, 89.75.Fb, 89.75.Hc}
\begin{abstract}
Clustering is well-known to play a prominent role in the description
and understanding of complex networks, and a large spectrum of tools and
ideas have been introduced to this end. In particular, it has been
recognized that the abundance of small subgraphs is important. 
Here, we study the arrangement of triangles in
a model for scale-free random graphs and determine the
asymptotic behavior of the
clustering coefficient, the average number of triangles, as well as
the number of triangles attached to the vertex of maximum degree. We
prove that triangles are power-law distributed among vertices and
characterized 
by both vertex and edge coagulation when the
degree exponent satisfies $2<\beta<2.5$; furthermore, a finite density
of triangles appears as $\beta=2+1/3$.
\end{abstract}
\maketitle
Graph representation is extensively used in many branches of science in order
to reduce the complexity of systems whose components have pairwise
interactions and where distance is irrelevant. One associates the components
of the system with the vertices of a graph and connects two of
them by an edge whenever a given property holds. It has turned out that
real-world networks, ranging from biology to physics, display common
topological features and, importantly, their degrees, 
power-law distributed (i.e., the number of vertices with $k$ edges
goes as $k^{-\beta}$ for some $\beta>2$, called the degree exponent),
reflect the presence of self-organizing phenomena underlying their
architecture~\cite{rev}. Owing to their power-law degree distribution
such networks are usually referred to as
scale-free networks~\cite{BA_science}, i.e., with no intrinsic
characteristic degree. 

A number of models aiming at understanding the features of
complex networks have been proposed, for instance 
\cite{Strogatz,prefer,Aiello,Newman_Strogatz,fitness} to cite a few.
In this work we focus on a model for power-law random graphs
\cite{VanVu0}  giving
good insight into the clustering properties.
We demonstrate that triangles coagulate 
into clusters and, in contrast to classical models for random
graphs (see~\cite{Bollobas} for
a review), they are power-law distributed: the probability for a
randomly selected vertex to participate in $t$ triangles goes as $\sim
t^{-(1+\beta)/2}$, with $\beta$ being the degree exponent. 
This scaling relation suggests that
triangles might be regarded as a fundamental
element for the characterization of real-world networks.

Our motivation resides in the recent attention 
devoted to the occurrence of small subgraphs, or motifs,
in scale-free networks. 
It has been observed~\cite{Alon_science,Alon_family} 
that some motifs are
over-represented in real-world networks as compared to randomized
networks with the same degree distribution.  
Usually the triangle is the building block of most motifs and
for random regular graphs it has been remarked~\cite{CollEck} that
when one imposes a finite density of triangles, they have the
tendency (i.e., higher probability) to organize themselves into 
complete subgraphs. Surprisingly, this phenomenon is more
likely when the imposed density of triangles is small. 

Our interest in triangles is also motivated by their interplay with a
simple transitivity relation and the fact that the clustering
coefficient can be used for breaking graphs up into
clusters carrying coherent information. The clustering coefficient
for a given vertex $i$ with degree $k_{i}$ is defined as~\cite{Strogatz}
$C_{i}=2t_{i}/(k^{2}_{i}-k_{i})$,
$t_{i}$ being the number of triangles attached to vertex $i$. Clusters
are obtained by fixing a threshold value 
and removing all vertices, and edges incident to them, 
with $C_{i}$ falling below it. This scheme was
applied to detect interest communities in the World Wide Web 
\cite{JPMoses} which turned out to be strongly affected 
by the presence of \textit{co-links}. This means that double edges 
with opposite direction are part
of a triangle with high probability, 
in line with findings in~\cite{Alon_family}, and thus emerge as
the basic unit of transitivity. A similar approach
has also been employed to organize lexical information into semantic
classes in order to differentiate meanings of 
ambiguous words~\cite{lexicon}.
Furthermore, related lines of research
\cite{Doro,Burda,Bianconi,CaldarelliLoops,Bianconi2,Ben,Bianconi3} 
have stressed the
importance and the abundance of cycles (or loops) in scale-free
networks.

\textit{The model.}
The best known model for random graphs is the Erd\"os-R\'enyi
model $\mathcal{G}(n,p)$ in which
every graph consists of $n$ vertices and each pair is connected by an
edge with uniform, independent probability $p$. The topology of such
graphs, however, shows marked deviations from that observed in real-world
networks. For instance, if $p=O(n^{-1})$ the degrees are Poisson
distributed, that is, the probability for a randomly
selected vertex to have $k$ edges is given by~\cite{Bollobas,Erdos}
$
P(k)=(\lambda^{k}/k!)e^{-\lambda},
$
where $\lambda$ is the average degree;
furthermore, triangles are almost surely (i.e., with probability equal
to one in the asymptotic limit) both edge and vertex disjoint.

Here, we investigate a generalization of the Erd\"os-R\'enyi model which
exhibits a power-law degree distribution. In our
analysis we shall follow closely Refs.~\cite{VanVu0}
to which we refer the reader for more details.

So, consider the set of random graphs $\mathcal{G}(\bm{w})$
in which every graph is specified by the average degree sequence
$\bm{w}=(w_{1},\dots,w_{n})$ arranged in decreasing order: $w_{1}\geq
w_{2}\geq\dots \geq w_{n}$.
Two vertices $i$ and $j$ are
connected with probability
$p_{ij}=w_{i} w_{j}/\sum_{l}w_{l}=\rho w_{i} w_{j}$,
where $1/\rho=\sum_{l=1}^{n}w_{l}$.
Importantly, by setting
\begin{equation}
w_{i}=c\ (i+i_{0})^{-1/(\beta-1)}
\label{eq:w}
\end{equation}
the number of vertices with degree $k$ turns out to be proportional 
to $k^{-\beta}$, and as a result the degrees are power-law distributed
with degree exponent $\beta$. 
The constants $c$ and $i_{0}$ appearing
in Eq.~(\ref{eq:w}) are determined by the average degree $d$ and
the maximum degree $m$. For $\beta>2$ one finds~\cite{VanVu0}
\[
c=d\ \frac{\beta-2}{\beta-1}\ n^{1/(\beta-1)}
\quad\text{and}\quad 
1+i_{0}=n\Big(\frac{d}{m}\frac{\beta-2}{\beta-1}\Big)^{\beta-1}\ .
\]
Probability normalization requires that
$m^{2}\leq \rho^{-1}$,
and so
$m\leq d^{1/2} n^{1/2}$.
In this model the average degree $d$ is a free parameter and in
the following we will assume that $d>1$; as a consequence, 
the maximum degree scales with $n$ as
\begin{equation}
m\sim n^{\alpha}\qquad\text{and}\qquad 0<\alpha\leq\frac{1}{2}\ .
\label{eq:m}
\end{equation}
Remark that $\alpha$ can be chosen independently of $\beta$.
Yet, another quantity of interest is the
second-order average degree $\tilde{d}=\rho\sum_{i}w_{i}^{2}$, 
in terms of which we
shall express most of our results. 
In the asymptotic limit we have~\cite{VanVu0}:
\begin{equation}
\label{eq:dtilde}
\tilde{d}=\left
\{
\begin{array}{ll}
d\frac{(\beta-2)^{2}}{(\beta-1)(3-\beta)}\Big
[\frac{m}{d}\frac{(\beta-1)}{(\beta-2)}\Big]^{3-\beta}\
&\text{if}\ 2<\beta<3\\
\frac{d}{2}\log\big(\frac{2m}{d}\big)
&\text{if}\ \beta=3\\
d\frac{(\beta-2)^{2}}{(\beta-1)(\beta-3)}
&\text{if}\ \beta>3
\end{array}
\right .
\end{equation}
making  apparent the existence of three different regimes as a
function of the degree exponent.

\textit{Results.}
The average number of triangles $t_{i}$ attached to vertex $i$ is
$
t_{i}
=\sum_{j>k\ j\neq i\ k\neq i}
p_{ij}p_{jk}p_{ki}\ .
$
This sum may be rearranged as
\begin{eqnarray}
\nonumber
t_{i}
&=&\frac{1}{2}\rho\ w_{i}^{2}\Big[
\rho^{2} \sum_{j,k} w_{j}^{2}w_{k}^{2}-
\rho^{2} \sum_{l}w_{l}^{4}\\
&&-2w_{i}^{2}\rho^{2}\sum_{l}w_{l}^{2}+2\rho^{2}w_{i}^{4}
\Big]\ .
\label{eq:ti}
\end{eqnarray}
In all regimes the leading term arises from the first
(double) sum 
in the right-hand side of the above expression. 
We find that $t_{i}/w_{i}^{2}=\rho\tilde{d}^{2}/2$ 
is of order $O(n^{-1}m^{2(3-\beta)})$
if $2<\beta<3$, of order $O(n^{-1}(\log n)^{2})$ 
if $\beta=3$, and of order $O(n^{-1})$ if
$\beta>3$. Neglected terms are at most of order $O(n^{-1}m^{3-\beta})$
if $2<\beta<3$, at most of order $O(n^{-1})$ if $\beta=3$, 
and at most of order $O(n^{-1}m^{3-\beta})$ if $\beta>3$
~\cite{t_more}. 
It readily follows that 
in the asymptotic limit the average
clustering coefficient of vertex $i$ reads
\[
C_{i}=\frac{2t_{i}}{w_{i}(w_{i}-1)}=
\frac{\rho(\tilde{d}w_{i})^{2}}{w_{i}(w_{i}-1)}
=\rho(\tilde{d})^{2}\big[1+O(w_{i}^{-1})\big]\ ,
\label{eq:C}
\]
and for sufficiently large values of $w_{i}$ this can be regarded as
independent of the degree of the anchor vertex. $C_{i}$ can be
interpreted as the probability that two neighbors of a vertex of
degree $w_{i}$ are joined together by an edge. By making use of
Eqs.~(\ref{eq:m}) and~(\ref{eq:dtilde}) one finds how the
clustering coefficient scales with the number of vertices $n$ in the
asymptotic limit. The average number of triangles attached 
to the vertex of maximum degree is simply given by 
$t_{1}=\rho (\tilde{d}m)^{2}/2$.
The results in the
asymptotic limit are summarized in Tab.~\ref{tab:Tab}.

The average number of triangles $T$ is obtained by calculating
\[
T=
\frac{1}{3}
\sum_{i}
t_{i}
=\frac{1}{3!}\big[
(\tilde{d})^{3}-3\tilde{d}\rho^{2}\sum_{l}w_{l}^{4}
+2\rho^{3}\sum_{l}w_{l}^{6}
\big]\ .
\]
As before, the dominant term arises from the first term in the
right-hand side of the above expression, $\tilde{d}^{3}/3!$, 
of order $O(m^{3(3-\beta)})$
if $2<\beta<3$, of order $O((\log n)^{3})$ if $\beta=3$, and of order 
$O(1)$ if $\beta>3$. The other ones are 
at most of order $O(m^{2(3-\beta)})$ if $2<\beta<3$, 
at most of order $O(\log n)$ if $\beta=3$, and
at most of order $O(m^{3-\beta})$ if $\beta>3$~\cite{T_more}. 
The asymptotic behavior of $T$ as a function 
of $m$ and $n$ for the different regimes
is also shown in Tab.~\ref{tab:Tab}.

We next address the question of how triangles are distributed
over the graph. Starting from~(\ref{eq:ti}) a simple calculation
proves that the probability for a randomly
selected vertex to participate in $t$ triangles goes as
\begin{equation}
P(t)\sim t^{-\delta}
\qquad\text{with}\quad\delta=\frac{1+\beta}{2}\ ,
\label{eq:Pt}
\end{equation}
and thus triangles are power-law distributed among vertices. 
\begin{table}
  \caption{Asymptotic behavior of the clustering coefficient, 
    $C$, the average number of triangles, 
    $T$, and the number of triangles attached to the
    vertex of maximum degree, $t_{1}$,  as a function of 
    the degree exponent $\beta$. 
    Recall that $m\sim n^{\alpha}$ with
    $0<\alpha\leq 1/2$.
    \label{tab:Tab}}
    \begin{ruledtabular}
    \begin{tabular}{llll}
	&
        $2<\beta<3$&
	$\beta=3$&
	$\beta>3$\\
        \hline\hline
	$C$&
	$\sim m^{2(3-\beta)}\ n^{-1}$&
	$\sim (\log m)^{2}\ n^{-1}$&
	$\sim n^{-1}$\\
	\hline
	$ T$&
	$\sim m^{3(3-\beta)}$&
	$\sim (\log m)^{3}$&
	$<\infty$\\
	\hline
	$t_{1}$&
	$\sim m^{2(4-\beta)}n^{-1}$&
	$\sim m^{2}(\log m)^{2}\ n^{-1}$&
	$m^{2}n^{-1}$\\
    \end{tabular}
    \end{ruledtabular}
\end{table}

\textit{Discussion.}
Some remarks on Tab. I are in order.
We see that irrespective of the choice of $\alpha$, 
Eq.~(\ref{eq:m}), the clustering coefficient remains a 
decreasing function of $n$ for $\beta>2$, that is always smaller than
1, and thus it preserves its probabilistic interpretation. 
The number of triangles always diverges
with $n$ in the range $2<\beta\leq 3$, corresponding to the regime observed
in real-world networks (see~\cite{rev} for examples); 
if instead $\beta >3$ then there are a finite number of triangles, as
in the Erd\"os-R\'enyi model. From Tab. I we can also see that
$\alpha=1/2$ seems to be a natural choice, and hence we set $\alpha$
equal to this value from here on.
 
Eq.~(\ref{eq:Pt}) is our main result.
This scaling relation tells us that with non-negligible probability some
vertices participate in a large number of triangles, which implies
that they are not scattered over the whole graph, as in the
Erd\"os-R\'enyi model, but coagulate around some vertices. 
Further understanding of such a phenomenon can be gained by
studying the inequality $t_{i} >w_{i}/2$, leading to
$i< O(1)
\times n(\tilde{d}^{2}/n)^{\beta-1}-i_{0}$.
Triangles start sharing a common edge when
$(n/i_{0})\times(\tilde{d}^{2}/n)^{\beta-1}$ 
is at least of order $O(1)$, that is,
for  $2<\beta<2.5$. Furthermore, the number of
vertices at which edge coagulation occurs goes as 
$n^{-(\beta-\beta_{-})(\beta-\beta_{+})}+O(n^{(3-\beta)/2})$ 
with $\beta_{\pm}=(3\pm\sqrt{5})/2$.

Note that as $\beta$ approaches $2$ 
the vertex of maximum degree sees  around itself a tightly connected
cloud since the clustering coefficient is close to being constant, 
whereas for $\beta\geq 3$ triangles are sparse in its neighborhood. 
In contrast, by looking at the fraction of triangles attached 
to it, that is
\begin{equation*}
\frac{t_{1}}{T}
\sim\left
\{
\begin{array}{ll}
n^{(\beta-3)/2}
&\text{if}\ 2<\beta<3\\
(\log n)^{-1}
&\text{if}\ \beta=3\\
<\infty
&\text{if}\ \beta>3
\end{array}
\right .
\end{equation*}
we deduce that triangles are spread over the graph for
$2<\beta<3$, and essentially centered around 
the vertex of maximum degree otherwise. 
Another quantity of interest is the density of triangles, namely
$T/n\sim n^{3(2-\beta)/2+1/2}$ for $2<\beta<3$,
and as $\beta=2+1/3$ we have a finite density of triangles.
\begin{figure}[hbtp]
\begin{center}
\includegraphics[width=8.5cm]{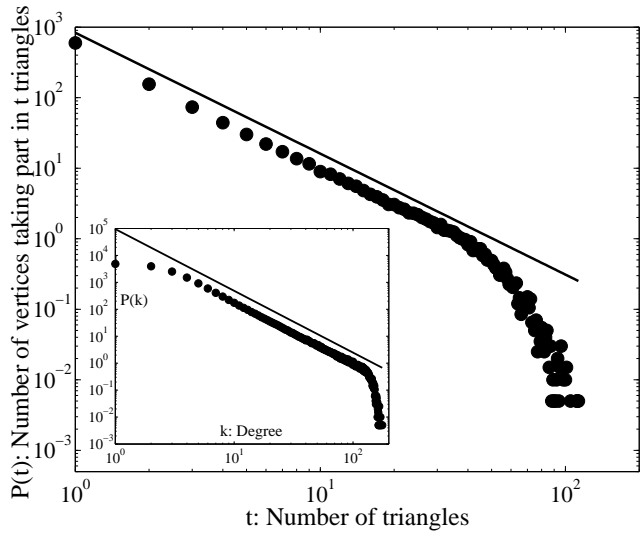}
\caption{The number of vertices participating
  in a given number of triangles as obtained from
  simulations for $\beta=2.2$. The number of vertices in graphs is set
  to $n=2\times 10^{4}$, the maximum degree to $m=\sqrt{n}$, and the
  average degree to $d=(\beta-1)/(\beta-2)$. Averages are taken over 200
  realizations and the scale of
  axes is logarithmic. The linear fit yields $\delta_{\rm{m}}=1.71\pm 0.02$. 
  For other values of $\beta$ the results are summarized
  in Tab.~\ref{tab:Tab2}. Inset: The degree distribution $P(k)$. The
  solid line has slope $-2.29\pm 0.02$.}
\label{fig:tri}
\end{center}
\end{figure}

\begin{figure}[hbtp]
\begin{center}
\includegraphics[width=8.5cm]{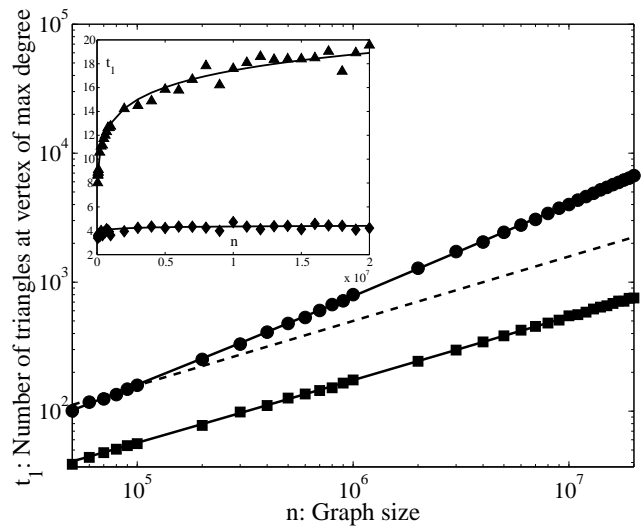}
\caption{The dependence of $t_{1}$, the average number 
  of triangles attached  
  to the vertex of maximum degree, on $n$, 
  the graph size, as obtained from
  simulations for $\beta=2.2$ (circles) and $2.5$ (squares). 
  Comparison is made with the theoretical prediction
  $t_{1}=\rho(\tilde{d}m)^{2}/2$ (solid lines). As before
  $d=(\beta-1)/(\beta-2)$ and $m=\sqrt{n}$; 
  averages are taken over 200 configurations and
  the scale of axes is logarithmic. The linear fit yields a slope of
  $0.71\pm 0.01$ for $\beta=2.2$ and of $0.48\pm 0.01$ for
  $\beta=2.5$; the theoretical value is given by $3-\beta$
  (cf. Tab.~\ref{tab:Tab}). The dashed line corresponds to 
  $t_{1}=\sqrt{n}/2$ marking the transition to edge
  coagulation. Notice that 
  for $\beta=2.5$ there is no edge coagulation, 
  whereas for $\beta=2.2$ there is; indeed, points obtained from
  simulations cross the dashed line.
  Inset: $t_{1}$ as a function of $n$ for
  $\beta=3$ (triangles) and $3.4$ (diamonds). Solid lines correspond
  to the theoretical predictions.}
\label{fig:size}
\end{center}
\end{figure}

Simulations have been performed in order to study the scaling
relation of Eq.~(\ref{eq:Pt}) as a function of $\beta$. 
Figure~\ref{fig:tri} illustrates the results for $\beta=2.2$.
Points obtained from simulations 
clearly follow a power-law with a cut-off as $t$ approaches
$t_{1}\approx 56$; the measured exponent
$\delta_{\rm{m}}$ is in accordance with the theoretical value. Table
\ref{tab:Tab2} shows the results for other values of $\beta$. Finite
size effects are more marked as $\beta$ approaches $3$. The reason
is that the number of triangles and, in particular, 
$t_{1}$, which determines the cut-off, increase with $n$ at a
slower rate (see  Tab.~\ref{tab:Tab}). In that respect 
it is worth noticing that for $\beta=2.2$ and $n=2\times 10^{4}$
vertices we have $m\approx 141$ and $t_{1}\approx 56$, and edge
coagulation does not occur since $t_{1}>m/2$ does not hold. This is
a finite size effect since for $n=10^{7}$ vertices we would have
$m\approx 3,162$ and $t_{1}\approx 4,033$, 
and the condition for edge coagulation is fulfilled. 
To make this point clearer we have investigated numerically 
$t_{1}$ as a function of $n$; the results are shown in 
Fig.~\ref{fig:size} and we see a good agreement between
simulations and theoretical predictions in the different regimes.
Obviously, the power-law behavior breaks down 
in the presence of a small, finite number of triangles 
on average, i.e. $\beta>3$. 
\begin{table}[htbp]
  \caption{The exponent characterizing the distribution of triangles among
  vertices, Eq.~(\ref{eq:Pt}), resulting from simulations as a function
  of the degree exponent $\beta$. Here $\delta_{\rm{m}}$ and
  $\delta_{\rm{t}}$ denote the measured and theoretical values, respectively.
    \label{tab:Tab2}}
    \begin{ruledtabular}
    \begin{tabular}{lcccc}   
      $\beta$&
      $2.2$&
      $2.3$&
      $2.5$&
      $2.8$
      \\
      \hline\hline
      $\delta_{\rm{m}}$&
      $1.71\pm 0.02$&
      $1.83\pm$ 0.04&
      $2.05\pm 0.05$&
      $2.5\pm 0.13$
      \\
      \hline
      $\delta_{\rm{t}}$&
      $1.6$&
      $1.65$&
      $1.75$&
      $1.9$
    \end{tabular}
    \end{ruledtabular}
\end{table}

We point out that the coagulation phenomenon reported in~\cite{CollEck} 
and the one investigated here are of quite a different
character. Specifically, in regular graph models the number of graphs
with a finite density of triangles is small and correspond 
statistically to graphs obtained by placing triangles so that to
construct the largest complete subgraph. Conversely, in power-law
random graphs it turns out that triangles are significant on 
average and display statistical regularities. The common feature
is that as topology departs from a certain degree of randomness it
gives rise to a pressure towards clustering and triangles arrange
themselves accordingly.

It is possible to make contact with models making use of
fitness variables. In Refs.~\cite{fitness,hidden} two vertices $i$ and $j$
are connected with probability $f(x_{i},x_{j})$, where $x_{i}$ and 
$x_{j}$ denote the intrinsic fitness of $i$ and $j$,
respectively. Fitness of vertices is distributed according to
$h(x)$. Within this model the number of triangles attached to a
vertex of fitness $x$ is
\begin{equation*}
t(x)=\frac{n^{2}}{2}\int_{0}^{\infty}f(x,y)f(y,z)f(z,x)h(y)h(z)
{\rm d} y{\rm d} z=\frac{n^{2}}{2}G(x)\ .
\end{equation*}
It follows that the probability for a randomly selected vertex to
participate in $t$ triangles can be written as
\begin{equation*}
P(t)=h\Big[G^{-1}\Big(\frac{2t}{n^{2}}\Big)\Big]
\frac{{\rm d}}{{\rm d}t}G^{-1}\Big(\frac{2t}{n^{2}}\Big)\ .
\end{equation*}
The statistical properties of graphs arise from the choice of $f$
and $h$ and one can prove that for a particular choice this model is
equivalent to the one studied here. We leave a detailed discussion
to a future publication.

A generalization of the model investigated here would consist in
implementing a non-trivial dependence of the clustering coefficient on
the degree. Note, however, that the mechanisms
responsible for clustering are basically the same, and 
in the case of a clustering
coefficient decreasing with the degree $k$ as $C\sim k^{-\gamma}$ 
we have
$P(t)\sim t^{-(1+\beta-\gamma)/(2-\gamma)}$.
We address the reader to Ref.~\cite{biology} for a study of the
presence of this scaling relation in biological networks. 
The purpose of~\cite{biology} was to establish a duality between 
large-scale topological organization and local subgraph structure in 
empirical networks. Our analysis differs from~\cite{biology} in that 
we have dealt with a probabilistic model allowing for a rigorous 
treatment of the asymptotic limit, but this is done at the 
expense of generality. Note that random growth processes have been
investigated within the framework of the same ideas in 
\cite{Alexei_growth}.

To summarize, in this work we have presented the study of a random graph
model and derived the asymptotic
behavior of some quantities describing the clustering properties,
coming to the conclusion that they are characterized by three
regimes, Tab.~\ref{tab:Tab}. The picture that emerges is that as the
degree exponent $\beta$ decreases the number of triangles increases
and arrange themselves into graphs so as to create tightly
connected cores around vertices of progressively smaller degree,
resulting in a power-law distribution, Eq.~(\ref{eq:Pt}). This is what
we refer to as coagulation of triangles. In itself, this
phenomenon dictates the abundance of recurring small patterns in the graph.

We thank A.-L. Barab\'asi, P. Collet, J.-P. Eckmann, 
E. Moses, A. V\'azquez and P. Wittwer for helpful
discussions. We are also grateful to the anonymous referees for their
comments on a previous version of this paper.
This work was supported by the Fonds National Suisse.


\begin{thebibliography}{2}
\bibitem{rev}R. Albert and A.-L. Barab\'asi, Rev. Mod. Phys.
  \textbf{74}, 47 (2002); M.E.J. Newman, SIAM Rev. \textbf{45}, 167
  (2003); S.N. Dorogovtsev and J.F.F. Mendes, Adv. Phys. \textbf{51},
  1079 (2002).

\bibitem{BA_science}A.-L. Barab\'asi and R. Albert, Science  
  \textbf{286}, 509 (1999).

\bibitem{Strogatz}D.J. Watts and H. Strogatz, Nature \textbf{393}, 440
  (1998). 

\bibitem{prefer}A.-L. Barab\'asi, R. Albert, and H. Jeong, Physica A
  \textbf{272}, 173 (1999). 

\bibitem{Aiello}W. Aiello, F. Chung, and L. Lu, \textit{Proceedings of
  the 32nd ACM Symposium on the Theory of Computing} (ACM, New York, 2000).

\bibitem{Newman_Strogatz} M.E.J. Newman, S.H. Strogatz, and D.J. Watts,
  Phys. Rev. E \textbf{64}, 26118 (2001).

\bibitem{fitness}G. Caldarelli, A. Capocci, P. DeLosRios, and
  M.A. Mu\~nuoz, Phys. Rev. Lett. \textbf{89}, 258702 (2002).

\bibitem{VanVu0}F. Chung, L. Lu, and V. Vu, Annals of Combinatorics
  \textbf{7}, 21 (2003); Proc. Natl. Acad. Sci.
  USA \textbf{100}, 6313 (2003).

\bibitem{Bollobas}B. Bollob\'as, \textit{Random Graphs} 
  (Academic Press, New York, 1985).

\bibitem{Alon_science}R. Milo, S. Shen-Orr, S. Itzkovitz, N. Kashtan,
  D. Chklovskii, and U. Alon, Science \textbf{298}, 824 (2002);
  S. Itzkovitz, R. Milo, N. Kashtan, G. Ziv, and
  U. Alon, Phys. Rev. E \textbf{68}, 26127 (2003).

\bibitem{Alon_family}R. Milo, S. Itzkovitz, N. Kashtan, R. Levitt,
  S. Shen-Orr, I. Ayzenshtat, M. Sheffer, and U. Alon, Science
  \textbf{303}, 1538 (2004).

\bibitem{CollEck}P. Collet and J.-P. Eckmann,
  J. Stat. Phys. \textbf{108}, 1107 (2002).

\bibitem{JPMoses}J.-P. Eckmann and E. Moses, Proc. Natl. Acad. Sci.
  USA \textbf{99}, 5825 (2002).

\bibitem{lexicon}B. Dorow, D. Widdows, K. Ling, J.-P. Eckmann, D. Sergi,
  and E. Moses, \textit{Proceedings of
  the 2nd Workshop Meaning05 on Developing Multilingual Web-Scale Language
  Technologies } (Meaning Project, Trento, 2005).

\bibitem{Doro}S.N. Dorogovtsev, Phys. Rev. E \textbf{69}, 27104
  (2004).
  
\bibitem{Burda}Z. Burda, J. Jurkiewicz, and A. Krzywicki, Phys. Rev. E
  \textbf{69}, 26106 (2004). 

\bibitem{Bianconi}G. Bianconi and A. Capocci,
  Phys. Rev. Lett. \textbf{90}, 78701 (2003).

\bibitem{CaldarelliLoops}G. Caldarelli, R. Pastor-Satorras, 
  and A. Vespignani, Eur. Phys. Jour. B \textbf{38}, 183 (2002). 

\bibitem{Bianconi2}G. Bianconi, G. Caldarelli, and A. Capocci,
  cond-mat/0310339. 

\bibitem{Ben}H.D. Rozenfeld, J.E. Kirk, E.M. Bollt, and
  D. Ben-Avraham, cond-mat/0403536.

\bibitem{Bianconi3}G. Bianconi and M. Marsilli, cond-mat/0502552.

\bibitem{Erdos}P. Erd\"os and A. R\'enyi,
  Publ. Math. Inst. Hungar. Acad. Sci. \textbf{5}, 17 (1960).

\bibitem{t_more} This last estimate is indeed valid in the interval
  $3<\beta<5$. For larger values of $\beta$ 
  we have: of order $O(n^{-2}\log n)$
  if $\beta=5$, and of order $O(n^{-2})$ if $\beta>5$. 

\bibitem{T_more} More precisely, this last estimate holds in the
  interval $3<\beta<5$. For $\alpha=1/2$, i.e. the case to which we
  shall specialize our analysis in the following, we have that
  neglected terms are of order $O(n^{-1}\log n)$ if $\beta=5$, and of
  order $O(n^{-1})$ if $\beta>5$.

\bibitem{hidden}M. Bogu\~na and R. Pastor-Satorras, Phys. Rev. E
  \textbf{68}, 36112 (2003).

\bibitem{biology}A. V\'azquez, R. Dobrin, D. Sergi, J.-P. Eckmann,
  Z.N. Oltvai, and A.-L. Barab\'asi, Proc. Natl. Acad. Sci. USA 
  \textbf{101}, 17940 (2004). 

\bibitem{Alexei_growth}A. V\'azquez, J. Oliveira, and
  A.-L. Barab\'asi, Phys. Rev. E \textbf{71}, 25103 (2005). 

\end{thebibliography}
\end{document}